\documentstyle[12pt]{article}
\begin{document}
\title{Constants of motion in the dynamics of a 2N-junction SQUID}
\author{C. Nappi$^a$, G.Filatrella$^{b,c}$, and S. Pagano$^a$\\
\noindent
a) Istituto di Cibernetica del CNR, via Toiano 6,\\ I-80072 Arco
Felice Italy\\
b) Department of Physics and Unit\'a INFM, University of Salerno,\\
I-84081 Baronissi (SA), Italy\\
c) Physikalisches Institut, Lehrstuhl Experimentalphysik II,\\
University of T\"ubingen, D-72076 T\"ubingen, Germany\\}
 \maketitle

\begin{center}
PACS number(s) 74.50+r, 85.25.Dq
\end{center}
\baselineskip=20pt
\begin{abstract}
We show that a 2N junction SQUID (Superconducting QUantum
Interference Device) made of 2N overdamped, shunted, identical junctions
may be described as a system having only 6 degrees of freedom for any $N\geq
3$.
This is achieved by means of the reduction
introduced by Watanabe and Strogatz (Physica D, {\bf 74}, (1994) 197)
for series biased arrays. In our case 6 rather than 3 degrees of freedom
are necessary to describe the system, due to the requirement of
phase quantization along the superconducting loop constituting the device.
Generalization to multijunction parallel arrays is straightforward.
\end{abstract} \baselineskip=22pt

\section{Introduction}
Two linear arrays, each containing N Josephson junctions, closed into a
superconductor loop constitute  a 2N-junction SQUID (Superconducting QUantum
Interference Device), (see Fig.1). Such multijunction structures are of
interest
in designing high gain SQUIDs{\bf\cite{darula94,Darula95,sarnelli95,suzuki93}}.
The case with a single
junction for each branch (N=1) is the well studied case of a
dc-SQUID{\bf \cite{barone82,likharev,tesche77}}.
Structures containing N=2, and N=3 junctions for each branch, have also been
fabricated and investigated in some detail{\bf\cite{darula94,sarnelli95}} .
Moreover theoretical analysis based on numerical simulations of the 2N-junction
SQUID has
appeared too{\bf\cite{suzuki93}}.
The aim of this
letter is to point out that, as for series arrays of overdamped
Josephson junctions shunted by passive elements, a 2N-junction SQUID,
made of overdamped (vanishing capacitance) identical junctions, exhibits
a highly degenerate dynamics. In particular we will show that the
transformation introduced by Watanabe and Strogatz (WS){\bf\cite{watanabe94}}
 may be fruitfully applied to the case considered
here, with few modifications, to demonstrate the existence of $2N-6$
constants of motion, {\em i.e.}, that any trajectory in the phase space
of a 2N SQUID is confined to a 6-dimensional subspace.
\indent
\section{The 2N-junction SQUID model}

\indent
The system we are interested in is sketched in Fig. 1 as a superconducting loop
in which each of the two branches include N identical Josephson junctions.
The behavior of the circuit is described by 2N dynamical
variables, $\{\phi^{(a)}_k\}$, $\{\phi^{(b)}_k\}$ ($k=1,...N$) representing the
gauge invariant phase difference across each junction for the branch $(a)$
and $(b)$, respectively. In the case of overdamped junctions, the governing
equations are {\bf \cite{barone82,likharev,tesche77}}
\begin{eqnarray}
\frac{\hbar}{2e}\frac{1}{R}\frac{d\phi^{(a)}_k}{dt} + I_0
\sin{\phi^{(a)}_k} - \frac{I}{2} &=& J\\
\frac{\hbar}{2e}\frac{1}{R}\frac{d\phi^{(b)}_k}{dt} + I_0
\sin{\phi^{(b)}_k} - \frac{I}{2} &=& -J
\end{eqnarray}
\noindent
$R$ is the normal resistance, $I_0$ is the critical current, $I$ is the
total bias current. The coupling terms  $J$ and $-J$ coincides with the
screening
current, which is determined through the fluxoid quantization:
\begin{equation}
J = \frac{\Phi_0}{4\pi L} \left \{ \sum_{i=1}^{N} \phi^{(a)}_i -
\sum_{i=1}^{N} \phi^{(b)}_i \right \} - \frac{\Phi_a}{2L}.
\end{equation}
\noindent
Here $\Phi_0$ is the flux quantum, $2L$ is the inductance of the loop,
and $\Phi_a$ is the externally applied magnetic flux.
The key observation is that the phases
$\phi^{(a)}_k$ and $\phi^{(b)}_k$ are globally coupled through Eq. (3)
and that for each subsystem, $(a)$ and $(b)$, this coupling is independent of
the index
$k$. This generalizes the case of globally coupled junction series arrays
studied by WS {\bf \cite{watanabe94}}. Indeed, although our system
is not globally coupled (there is a change of sign of the coupling term when
passing
from a branch to another), it can be split into two globally coupled
sub-systems
. Each subsystem contains N series connected junctions and satisfies the
conditions for the applicability of the reduction introduced by WS.
If we introduce  the dimensionless
variables
\begin{eqnarray}
\nonumber
t^{\prime} = \frac{2\pi I_0 R}{\Phi_0}t \\
\nonumber
i = \frac{I}{I_0} \\
\nonumber
j = \frac{J}{I_0} \\
\nonumber
\beta = \frac{2LI_0}{\Phi_0} \\
\nonumber
{\Phi_a}^{\prime} = \frac{\Phi_a}{\Phi_0}
\end{eqnarray}
\noindent
 Eq.s (1-3) take the form (dots denote derivative with
respect to the normalized time):
\begin{eqnarray}
\dot{\phi}_k^{(a)} &=& \frac{i}{2} - j - sin{\phi_k^{(a)}}; \,\,\,
,k=1,...,N\\
\dot{\phi}_k^{(b)} &=& \frac{i}{2} + j - sin{\phi_k^{(b)}}; \,\,\,
,k=1,...,N\\
j &=& \frac{ \sum_{k=1}^{N} \phi^{(a)}_k - \sum_{k=1}^{N} \phi^{(b)}_k
-2 \pi {\Phi_a}^{\prime}}{\pi \beta} .
\end{eqnarray}
\noindent
 Therefore we introduce a change of coordinates separately for each
 branch, as{\bf \cite{watanabe94}} :
\begin{eqnarray}
\tan \left [\frac{1}{2}\left (\phi_k^{(a)}(t^{\prime}) -
{\Theta}^{(a)}(t^{\prime}) \right
) \right ] = \sqrt{
\frac{1+\gamma^{(a)}(t^{\prime})}{1-\gamma^{(a)}(t^{\prime})}}
\tan \left [\frac{1}{2}\left (
\psi_k^{(a)} -\Psi^{(a)}(t^{\prime})\right ) \right ] \,\,\, k=1,...N\\
\tan \left [\frac{1}{2}\left (\phi_k^{(b)}(t^{\prime}) -
{\Theta}^{(b)}(t^{\prime}) \right
) \right ] = \sqrt{
\frac{1+\gamma^{(b)}(t^{\prime})}{1-\gamma^{(b)}(t^{\prime})}}
\tan \left [\frac{1}{2}\left (
\psi_k^{(b)} -\Psi^{(b)}(t^{\prime})\right ) \right ] \,\,\, k=1,...N.
\end{eqnarray}
\noindent
Here ${\Theta^{(a)}}$,$\gamma^{(a)}$,$\Psi^{(a)}$,${\Theta^{(b)}}$,
$\gamma^{(b)}$, and $\Psi^{(b)}$, are unknown functions of time
($0\leq\gamma^{(a)},\gamma^{(b)}<1$),
$\psi_k^{(a)}$
and $\psi_k^{(b)}$ are $2N$ constants.
It is a straightforward exercise to prove that the transformations (7,8)
reduce the 2N Eq.s (4-6), for every $N \ge 3$, to the following six
equations:
\begin{eqnarray}
\dot{\gamma}^{(a)} &=& -(1-\gamma^{(a) 2}) \cos {\Theta}^{(a)},\\
\gamma^{(a)}\dot{\Psi}^{(a)} &=& \sqrt{1-\gamma^{(a) 2}}\sin
{\Theta}^{(a)},\\
\gamma^{(a)}\dot{{\Theta}}^{(a)} &=& \gamma^{(a)} \left (\frac{i}{2}
-  j \right ) + \sin {\Theta}^{(a)},\\
\dot{\gamma}^{(b)} &=& -(1-\gamma^{(b) 2}) \cos {\Theta}^{(b)},\\
\gamma^{(b)}\dot{\Psi}^{(b)} &=& \sqrt{1-\gamma^{(b) 2}}\sin
{\Theta}^{(b)},\\
\gamma^{(b)}\dot{{\Theta}}^{(b)} &=& \gamma^{(b)} \left (\frac{i}{2}
+  j \right ) + \sin {\Theta}^{(b)}.
\end{eqnarray}
\noindent
Eq.s (9-14) form a closed system for the six unknown functions, since $j$
can be expressed in terms of the new variables
inserting the change of coordinates (7,8) in Eq. (6).
The 'frozen phases' (adopting the WS's language) appear in the system
merely as parameters. Some comments are in order. Firstly, the above system
becomes
singular if either $\gamma^{(a)}$, or $\gamma^{(b)}$ vanish.
Moreover negative values for  $\gamma^{(a)}$, or $\gamma^{(b)}$ seem not
excluded,
contrary to their definition.
WS have shown that these troubles are not essential, but just an artifact of
the
chosen coordinate system. As a matter of fact they can be avoided by the
additional change of coordinates:
\begin{eqnarray}
x^{(a)} &=& \gamma^{(a)} \cos {\Theta}^{(a)},\\
y^{(a)} &=& \gamma^{(a)} \sin {\Theta}^{(a)},\\
z^{(a)} &=& \Theta^{(a)} - \Psi^{(a)},\\
x^{(b)} &=& \gamma^{(b)} \cos {\Theta}^{(b)},\\
y^{(b)} &=& \gamma^{(b)} \sin {\Theta}^{(b)},\\
z^{(b)} &=& \Theta^{(b)} - \Psi^{(b)}
\end{eqnarray}
Thus the system, in terms of the new variables, becomes :
\begin{eqnarray}
\dot{x}^{(a)}  &=&  -1 + x^{(a) 2} -y^{(a)} \left (\frac{i} {2} - j \right ),
\\
\dot{y}^{(a)}  &=&  x^{(a)} y^{(a)} + x^{(a)} \left (\frac{i} {2} - j \right ),
\\
\dot{z}^{(a)}  &=&  \left (\frac{i} {2} - j \right ) + \frac {1 -
\sqrt{1-x^{(a)
2}-y^{(a) 2}}} {x^{(a) 2}+y^{(a) 2}} y^{(a)} ,\\
\dot{x}^{(b)}  &=&  -1 + x^{(b) 2} -y^{(b)} \left (\frac{i} {2} + j \right ),
\\
\dot{y}^{(b)}  &=&  x^{(b)} y^{(b)} + x^{(b)} \left (\frac{i} {2} + j \right ),
\\
\dot{z}^{(b)}  &=&  \left (\frac{i} {2} + j \right ) + \frac {1 -
\sqrt{1-x^{(b)
2}-y^{(b) 2}}} {x^{(b) 2}+y^{(b) 2}} y^{(b)}
\end{eqnarray}
which is well-behavied even at $\gamma^{(a)} = 0$, or $\gamma^{(b)} = 0$
The second point is about the initial conditions. Given 2N arbitrary initial
junction
phase values, the values for the $2N$ 'frozen phases' $\psi_k^{(a),(b)}$ and
the
initial values for the six 'reduced variables'
$x^{(a),(b)},y^{(a),(b)},z^{(a),(b)}$
should be obtained from Eqs. (7,8). This of course can be done in many ways: a
simple
and natural choice is the 'identity transformation' :
\begin{equation}
x^{(a)}(0) = y^{(a)}(0) = z^{(a)}(0) = x^{(b)}(0) = y^{(b)}(0) = z^{(b)}(0) = 0
\end{equation}
which straightforwardly gives :
\begin{equation}
\psi_k^{(a),(b)} = \phi_k^{(a),(b)}(0)
\end{equation}
The investigation of the meaning of such moltiplicity of choice
for the initial values is an important issue for the understanding of the
dynamics
 of the system, but goes outside the scope of this letter.We just comment that,
as (27) and (28) show, there is at least one way to effectively achieve the
reduction.

We have numerically verified, for few values of the parameters, that the
reduced system of six equations yelds the same results of the originary system,
Eqs.(4-6). This was checked with increasing number of junctions up to N= 20
junctions per branch. Care must be taken in reconstructing the phase values
from
the six unknown functions to avoid spurious $2\pi$ jumps in the solution.

\section{Parallel arrays}
The above property may be exploited, more generally, to a multijunction
parallel array  of $M$ branches each branch containing $N$ identical junctions
(Fig. 2).
Then Eq.s (1-2) can be generalized as follows:
\begin{eqnarray}
\frac{\hbar}{2e}\frac{1}{R}\frac{d\phi^{(m)}_k}{dt} + I_0
\sin{\phi^{(m)}_k}= I^{(m)} \;\;\; (m=1,2,...,M; \; k = 1,2,...,N)
\end{eqnarray}
\noindent
where $I^{(m)}$ is the current in the $m-th$ branch, and $\phi^{(m)}_k$ is
the phase of the k-th junction in the m-th branch.
The $I^{(m)}$ sum, of course, to the total bias current $I$ :
\begin{equation}
\sum_{l=1}^{M} I^{(l)}   =  I
\end{equation}
Moreover the condition for the fluxoid quantization in each loop writes :
\begin{equation}
\sum_{i=1}^{N} \phi_i^{(j)} - \sum_{i=1}^{N}\phi_i^{(j+1)} = 2 \pi \frac
{\Phi_a} {\Phi_0}
+ \frac {2 \pi} {\Phi_0} \sum_{l=1}^{M} c_l^{(j)} I^{(l)} \\
\nonumber
\,\,\,\,\,\,\,\,\,j=1,\ldots,M-1
\end{equation}
Here $c_l^{(j)}$ is the mutual inductance coefficient accounting for the
contribution
to the magnetic flux at the loop $(j,j+1)$due to the current
flowing in the $l$-th branch.
These coefficients are determined by the specific geometry of the circuit.
Eqs. (30) and (31) are a set of linear equations for the unknown $I^{(m)}$, and
allow
a unique determination of their value as function of all the phase variables.
In the simple case in which only the nearest neighbours branch currents
contribute
to the magnetic flux in each loop{\bf \cite{phillips93,reinel94}}, $c_l^{(j)}$
assume the simple form :
\begin{equation}
c_l^{(j)} = L ( \delta_l^{(j+1)} - \delta_l^{(j)})
\end{equation}
where $L$ is the inductance of each branch and $\delta_l^{(j)}$ is the
Kronecker's symbol.
Under these simplifying hypotheses it is easy to show that :
\begin{equation}
I^{(m)} = \frac {I} {M} - \frac {\Phi_a} {L} \left ( \frac {M+1} {2} -m \right
)
- \frac {\Phi_0} {2 \pi L} \sum_{i=1}^{N} \phi_i^{(m)}
+ \frac {\Phi_0} {2 \pi M L} \sum_{k=1}^{M} \sum_{i=1}^{N} \phi_i^{(k)}
\nonumber
\,\,\,\,\,\,\,\,\,m=1,\ldots,M
\end{equation}

The $N\times M$ Eqs. (29) and (33) describe completely a multi-junction
parallel
array,
and, as in the case of the 2Nj-SQUID, it can be viewed as M sub-systems each
one globally coupled to the rest of the system. The voltage V developed
across the circuit  may be obtained by time derivating Eq.(33):
\begin{equation}
V \equiv \frac {\hbar}
{2e}\sum_{i=1}^{N}\frac{d\phi^{(m)}_i}{dt}+L\frac{dI^{(m)}}{dt}
=\frac {\hbar} {2eM} \sum_{k=1}^{M} \sum_{i=1}^{N} \frac{d\phi^{(k)}_i}{dt}
\end{equation}
It should be pointed out that the simplification introduced here (to neglect
off diagonal terms of the mutual inductance matrix) is not essential to
obtain a global coupling, {\em i.e.} also the more general case will lead
qualitatively to the same kind of (global) coupling. As in the previous
discussion
of the 2N-junction SQUID, it is therefore possible to introduce $M$
transformations analogous to  (7,8)
(one for each superconducting branch) and to conclude that only the following
$3M$ equations $(m=1...M)$

\begin{eqnarray}
\dot{x}^{(m)}  &=&  -1 + x^{(m) 2} -y^{(m)}\left (\frac {I^{(m)}} {I_o}\right
),
\\
\dot{y}^{(m)}  &=&  x^{(m)} y^{(m)} + x^{(m)}\left (\frac {I^{(m)}} {I_o}\right
), \\
\dot{z}^{(m)}  &=&  \frac {I^{(m)}} {I_o} + \frac {1 - \sqrt{1-x^{(m) 2}-y^{(m)
2}}} {x^{(m) 2}+y^{(m) 2}} y^{(m)}
\end{eqnarray}

should actually be solved for the $3M$ variables
$\{ x^{(m)},y^{(m)},z^{(m)}\}$, rather than $N\times M$. It should finally
remarked
that the  all identical junction assumption may be partially relaxed.
Indeed no particular troubles are met in the previous derivation if one assumes
 the junctions  in a branch (or even their number) to be different from
  those of another branch. That is to say, the requirement of identical
parameters
   concerns only  the junctions in the same branch.

\section{Conclusion}
The possibility to reduce an $N \times M$ system to $3 \times M$ one is
in itself of some advantage to simplify numerical simulations. Still we
believe that the major achievement that we have reached is that the
low dimensional motion for 'globally coupled' systems  with sine nonlinearity
can be extended also to sets of equations that are not 'globally coupled' in a
strict sense, but can be divided in subsystems exhibiting global coupling (at
the price to increase the number of dynamical variables necessary to describe
the system). So we have explicitly proved that phase quantization in
superconducting loops has no other effect that to introduce 3 more
variables for each loop. Beside the extension of the theory there is
also a consequence of practical importance: neutral stability is a
drawback for practical applications (as local oscillators) because of
the consequent deterioration of the linewidth of the emitted microwave.
Our findings can be interpreted as a criterion for the maximum
number of junctions that can be inserted in a loop: above $3$ the system
undergoes a reduction similar to that illustrated in this letter. It
should be noted that this criterion is not a positive one, {\em i.e.},
there might be configuration that are neutrally stable but do not
contain more than $3$ junctions per loop~{\bf
\cite{tsang91,filatrella95}}.
At this stage it is not known to which extent the formal analogy between the
system studied in Ref. {\bf \cite{watanabe94}} may be carried on. Indeed, there
are differences in the actual form of the coupling term that will lead to
substantial differences: for instance Eq.s (4-6) do not exhibit the important
property of {\em reversibility} (changing $\phi^{(m)}_k$ in $-\phi^{(m)}_k$
and $t^{\prime}$ in $-t^{\prime}$ does not leave unchanged the equations).
 This will lead to profound effects on the
dynamics of the system.

\section{Acknowledgement}

\indent
We wish to thank R. D. Parmentier for useful suggestions. The work was
partially
supported by the Progetto Finalizzato "Superconductive
and Cryogenic Technologies" of the Italian National Research Council and the
EU-Science Project no. SC1-CT91-0760 "Coupled Josephson junctions".
GF wishes to thank EU for financial support  under the ESPRIT project No. 7100.
His stay in Germany was made possible through the ``Human Capital and Mobility
program'', contract No. ERBCHRXCT 920068.
\medskip
\par

\section*{Figure Captions}
\begin{itemize}
\item[Fig.1]  Schematic circuit model for a 2N-junction SQUID;
\item[Fig.2] Schematic circuit model for a multi-junction parallel array;
\end{itemize}

\end{document}